%% file: main.tex
\documentclass[runningheads]{llncs}
\usepackage{graphicx}
\usepackage{amsmath}
\usepackage{amssymb}
\usepackage{multirow}
\usepackage[ruled,vlined]{algorithm2e}
\usepackage{wrapfig}
\usepackage[font=small]{caption}
\usepackage{xcolor}

\usepackage{hyperref}
\begin{document}
\title{Learning Guided Electron Microscopy\\with Active Acquisition}
%
%
\author{Lu Mi\inst{1}, Hao Wang\inst{1}, Yaron Meirovitch\inst{1,2}, Richard Schalek\inst{2}, Srinivas C. Turaga\inst{3}, Jeff W. Lichtman\inst{2}, Aravinthan D.T. Samuel\inst{2}, Nir Shavit\inst{1}}

\authorrunning{L. Mi et al.}
\institute{Massachusetts Institute of Technology, MA, USA\\
\and
Harvard University, MA, USA\\
\and 
HHMI Janelia Research Campus, VA, USA\\
\email{lumi@mit.edu}}

\maketitle  
\input{source/abstract}
\input{source/introduction}

\input{source/methodology}

\input{source/results}
\input{source/performance}

\input{source/conclusion}
\input{source/Acknowledgement}

%
%
\bibliographystyle{splncs04}
\bibliography{reference/reference}

\input{source/supp}

\end{document}

%% file: source/abstract.tex
\begin{abstract}

Single-beam scanning electron microscopes (SEM) are widely used to acquire massive data sets for biomedical study, material analysis, and fabrication inspection. Datasets are typically acquired with uniform acquisition: applying the electron beam with the same power and duration to all image pixels, even if there is great variety in the pixels' importance for eventual use. Many SEMs are now able to move the beam to any pixel in the field of view without delay, enabling them, in principle, to invest their time budget more effectively with non-uniform imaging.    

In this paper, we show how to use deep learning to accelerate and optimize single-beam SEM acquisition of images. Our algorithm rapidly collects an information-lossy image (e.g. low resolution) and then applies a novel learning method to identify a small subset
of pixels to be collected at higher resolution based on a trade-off between the saliency and spatial diversity. We demonstrate the efficacy of this novel technique for active acquisition by speeding up the task of collecting connectomic datasets for neurobiology by up to an order of magnitude. Code is available at~\url{https://github.com/lumi9587/learning-guided-SEM}.

\keywords{Electron Microscope \and Active Acquisition \and Determinantal Point Process.}

\end{abstract}

%% file: source/introduction.tex
\section{Introduction}
\begin{figure}[t]
\begin{center}
    \includegraphics[width=1\textwidth]{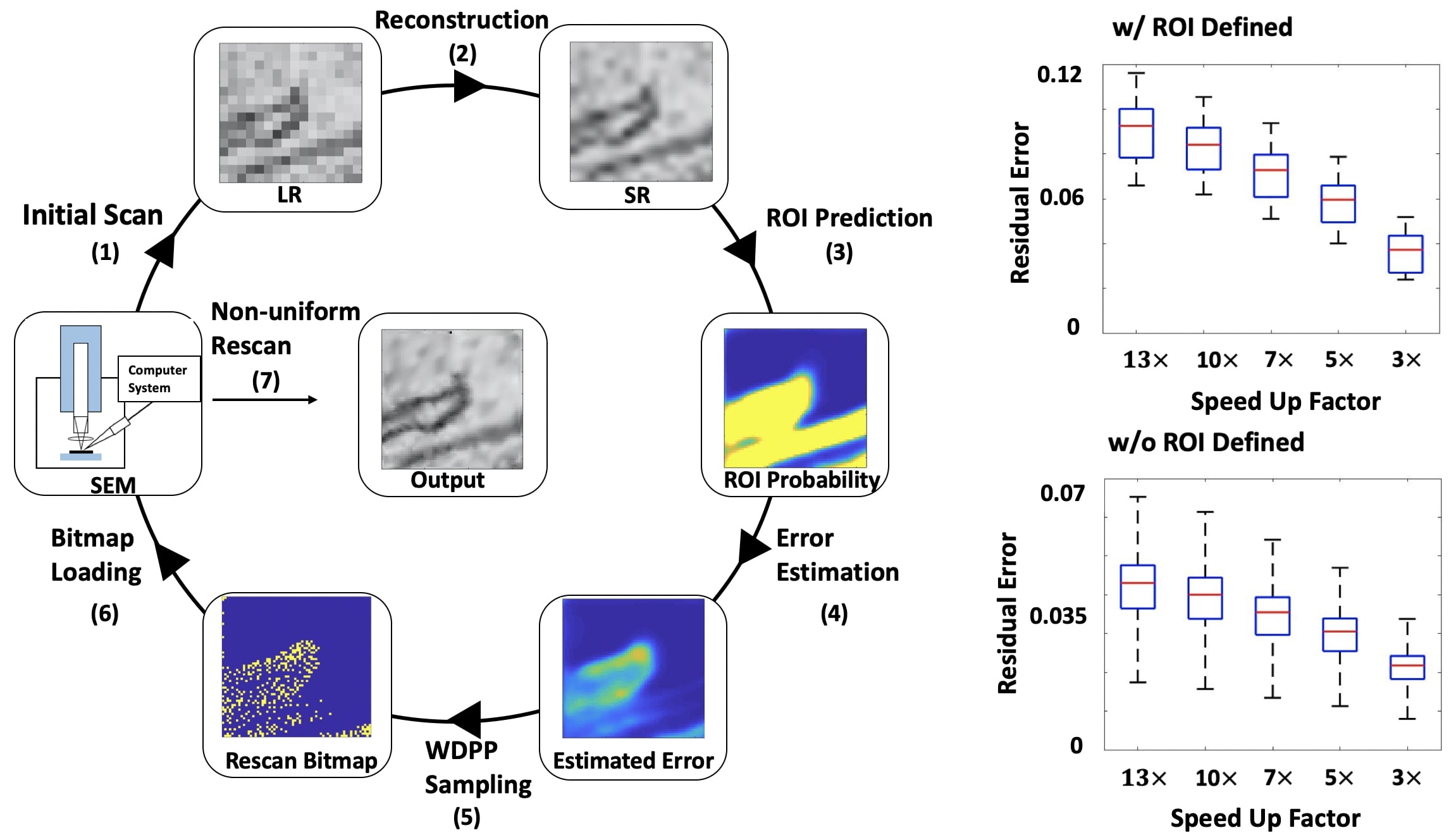}
\end{center}
    \caption{\textbf{Left}: Overview of our learning guided electron microscope with ROI defined. The SEM firstly performs the initial scan to collect the low-resolution image $I_{LR}$ (1). The reconstruction step is applied to generate a super-resolved image $I_{SR}$ (2). Then the pixel of interest (3) and estimated error (4) is identified. The next step is to implement diversified sampling based on estimated error using weighted determinantal point process (WDPP) (5). Finally a sparse bitmap is sent to SEM (6) to perform rescan (7). See more details in Section~\ref{sec:pipeline}. \textbf{Right}: The box plots represent different speedup factors with \textcolor{black}{residual error ($L_1$ loss)} using our active acquisition pipeline on \emph{SNEMI3D}.} 
    \label{fig:pipeline} 
\end{figure}
Scanning electron microscopes are widely used for nanometer-scale imaging in diverse applications including  structural biology \cite{kasthuri2015saturated,helmstaedter2011high}, materials analysis~\cite{pandey2003material}, and semiconductor fabrication \cite{newell2016detection}. In most cases, an electron beam is applied with the same power and duration to all image pixels. 
This is fundamentally inefficient since the saliency of each pixel might be heterogeneous. 

In this paper, we design an imaging strategy for a scanning electron microscope (SEM) that mimics the human visual system. The human visual system quickly decides saliency by first using a low-resolution (non-foveal) collection and then applies the high-resolution fovea to dwell on important parts \cite{thorpe1996speed}. Most scanning electron microscopes are ideally suited for such active and adaptive image acquisition; their scan generators can instantaneously deflect the beam to any arbitrary position and then stably dwell to acquire each pixel \cite{anderson2013sparse,semreview}. Because of fast beam re-positioning, an arbitrary distribution of pixel locations has negligible impact on scanning time. 
Therefore, for an image where only 5\% of pixels need scanning, one can ideally speed up its acquisition by 20 folds~\cite{anderson2013sparse,potocek2020sparse}. 

Most images are characterized by two types of saliency: \textit{interest} 
and \textit{information density}.
Conventional SEM ignores both types of saliency and scans the whole image using identical resolution with low efficiency~\cite{flegler1997scanning}. 
In this work, we develop a new method to accelerate and optimize SEM acquisition. To do this, we have built an active acquisition algorithm that intelligently budgets the operation of a scanner to focus on regions \textcolor{black}{with high saliency (either high interest or high information density)} within an image. 

We apply our technology to one critical area using SEM in connectomics, which aims for the reconstruction of synaptic connectivity maps for brain tissue~\cite{lichtman2014big,kasthuri2015saturated}. 
So far, connectomics has only been applied to a small number of specimens because of the enormous burden in acquiring and analyzing datasets that can easily span terabytes and petabytes \cite{jarrell2012connectome,yan2017network}. For example, scanning a cubic millimeter of brain tissue at the resolution needed for connectomics ($4nm \times 4nm \times 30nm$ per voxel) requires two thousand trillion voxels (2PBs of data). A typical SEM running at one million voxels per second would require 63 years to do this. 

Here, we design to accelerate SEM image acquisition for connectomics by exploiting the sparsity of salient pixels in images. The essential goal in connectomics is to map neural circuitry~\cite{46992,meirovitch2019cross}. The only structures that contain pixels with high interest are membrane of neurons, and objects associated with synapses. All other intracellular objects and extracellular space have low saliency. For typical mammalian tissues, neural membranes account for only 5-10\% of the images and synapses account for even less~\cite{kasthuri2015saturated}. This sparsity of salient pixels suggests significant potential for speeding up acquisition. \textcolor{black}{In this work, without loss of generality, we define the membrane boundary of individual neurons as the region of interest (ROI), to demonstrate the effectiveness of our pipeline.} 

In contrast to previous work using image reconstruction~\cite{eldar2012compressed,gan2007block,ledig2017photo,isola2017image,weigert2018content,wang2019deep,fang2019deep,buchholz2019content}, or multi-beam approaches that use highly parallelized but expensive microscope systems~\cite{eberle2015high}, our work aims to guide widely available single-beam SEMs to collect salient image pixels, thereby reconstructing essential regions at high resolution, as shown in Fig.~\ref{fig:pipeline}. 
Our major contributions are:

\begin{itemize}

\item We are the first, to our knowledge, to cast the acquisition of electron microscopes as a learning-guided sampling problem and thereby capable to achieve significant speedup.

\item We present an effective and principled sampling technique, weighted determinantal point process (WDPP), that optimizes pixel selection based on their saliency and spatial diversity. 

\item We present a new active-acquisition pipeline for SEM that executes non-uniform pixel-wise scanning, and demonstrate a potential speedup rate of up to an order of magnitude on real-world connectomic datasets.

\end{itemize}

%% file: source/methodology.tex
\section{Methodology}
In this paper, we formulate a learning-guided sampling problem to speed up SEM acquisition. The goal is to intelligently sample a subset of pixels in a way that balances the following trade-off: 
\begin{align*}
\mathcal{C} = \underbrace{|I_{HR} - R(I_{HR} \odot B)|}_{reconstruction\ loss} + \lambda \underbrace{\sum_{i,j}{B^{(i,j)}}}_{acquisition\ cost}.
\end{align*}

Here $I_{HR}$ denotes a high-resolution image, $B$ is the bitmap with the superscript indexing positions to indicate locations of sampled pixels, $R(\cdot)$ is the reconstruction function, $\odot$ is the Hadamard (element-wise) product, and $\lambda$ is a hyperparameter balancing the trade-off between reconstruction loss and acquisition cost. Unlike other works~\cite{ledig2017photo,isola2017image,weigert2018content,wang2019deep} focusing on improving $R(\cdot)$ given a low-resolution image $I_{LR} =  I_{HR} \odot B$, our work assumes a fixed $R(\cdot)$ and instead tries to find a reasonable bitmap $B$ that can achieve low reconstruction loss with low acquisition cost. We do this via the proposed WDPP sampling technique, which selects pixels based on spatial diversity as well as saliency (quantified as estimated error). In the following, we describe the overview of our active acquisition pipeline, introduce its key components (i.e., binarized error estimation in Section~\ref{section: error estimation} and WDPP in Section~\ref{section: DPP Sampling}) as well as our technical contributions.

\subsection{Active Acquisition Pipeline}
\label{sec:pipeline}

Below we describe individual steps of the active acquisition pipeline in Fig.~\ref{fig:pipeline}. 

\textbf{Initial Scan, Reconstruction, and ROI Prediction}: As the first step, SEM performs the \emph{initial scan} of a low-resolution image $I_{LR}$ with negligible cost. 
The next step is to apply a \emph{reconstruction} model $R$ to $I_{LR}$ to generate super-resolved image $I_{SR} = R(I_{LR})$. The model can use either learning-based reconstruction methods or simple interpolation rules such as bicubic. The third step (\emph{ROI Prediction}) is using ROI detector $F_{ROI}$ to predict the saliency score for each pixel. Note that we are interested in two types of tasks in this paper: tasks with and tasks without ROI defined. \emph{ROI Prediction} is not applied for the task without ROI defined. This task only considers regions with high information density as saliency. These steps correspond to (1)-(3) in Fig.~\ref{fig:pipeline}.

\textbf{Binarized Error Estimation}: 
The fourth step, shown as (4) of Fig.~\ref{fig:pipeline}, is to \emph{estimate the prediction error}. For the tasks without ROI defined, the ground truth error is defined as $L_1$ loss $ \mid I_{HR}-I_{SR} \mid$; for the tasks with ROI defined, the ground truth error is defined as $L_1$ loss $ \mid F_{ROI}(I_{HR})-F_{ROI}(I_{SR}) \mid$. These errors will be estimated through an efficient and simple learning based method we propose in Section~\ref{section: error estimation}. 

\textbf{Diversified Sampling}: This step, as shown in (5) of Fig.~\ref{fig:pipeline}, is to perform WDPP sampling based on the estimated error map from the previous step. The goal is to select \textcolor{black}{$K$} samples contributing the largest estimated error while balancing the spatial diversity at the same time. Details are in Section~\ref{section: DPP Sampling}.

\textbf{Bitmap Loading and Rescan}: Once the locations of sampled pixels (produced by WDPP) are encoded into bitmaps and loaded into the SEM, the SEM will perform rescan based on the sparse bitmap. The final output $I_{OUT}$ is then reconstructed with recollected pixels during rescan as well as pixels in $I_{LR}$ collected in the initial scan.

\subsection{Binarized Error Estimation}
\label{section: error estimation}

\begin{figure*}[t]

\begin{center}
    \includegraphics[width=1\textwidth]{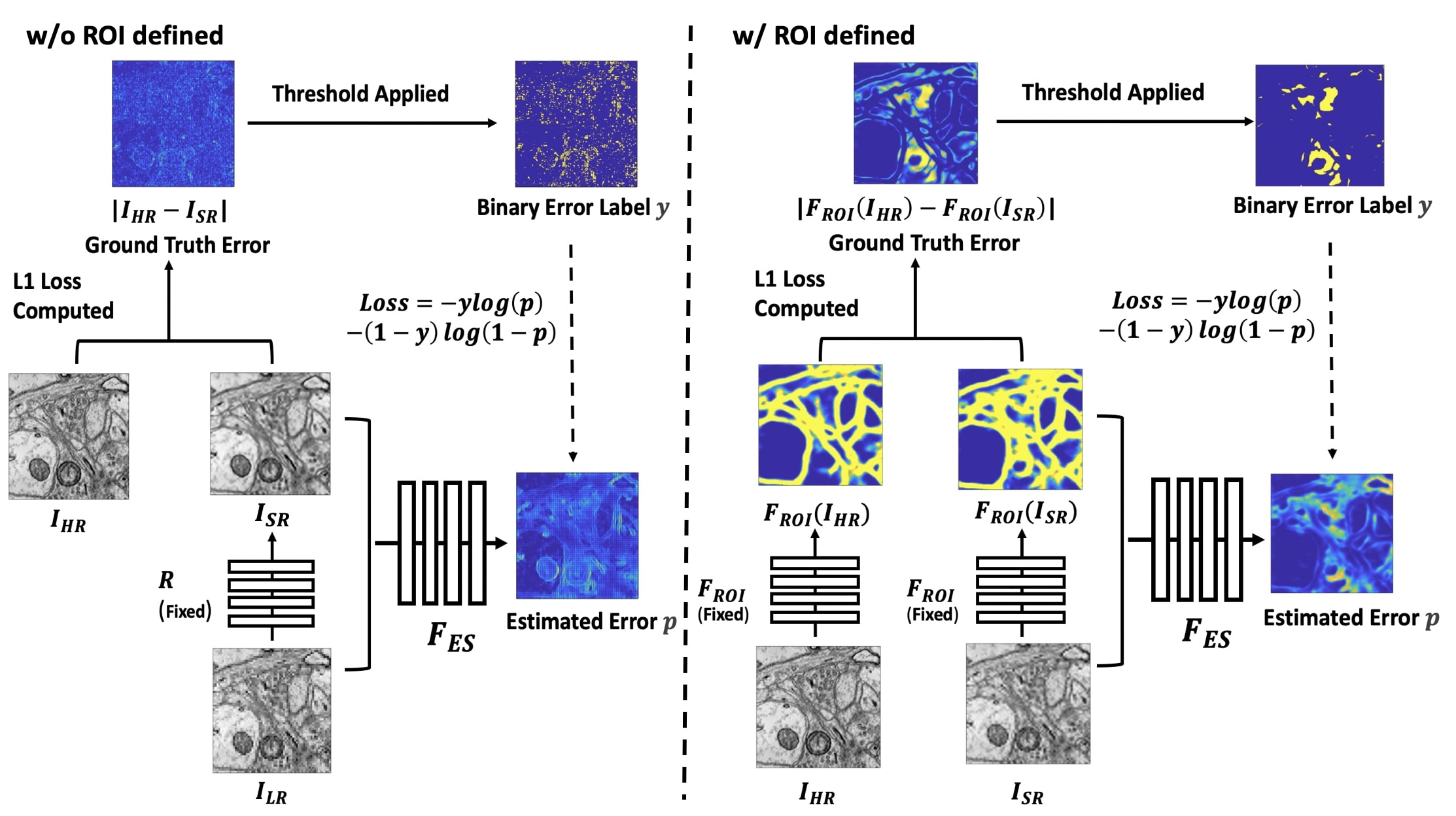}
\end{center}
    \caption{The overview of our end-to-end learning framework to train the error estimation networks $F_{ES}$. \textcolor{black}{For the task without ROI defined (\textbf{left}), the inputs of $F_{ES}$ are concatenation of $I_{SR}$ and $I_{LR}$. For the task with ROI defined (\textbf{right}), the inputs of $F_{ES}$ are concatenation of   $F_{ROI}(I_{SR})$ and $I_{SR}$.} The regression task is reformulated as a binary classification problem after applying a threshold to the ground truth error. ROI detector $F_{ROI}$ and reconstruction model $R$ are fixed \textcolor{black}{during the training of $F_{ES}$.}}
    \label{fig:error_etsimation_framework} 
\end{figure*}

One key component of the pipeline is our proposed \emph{binarized error estimation}. It is a simple and efficient supervised learning method to estimate pixel-wise error. 

As shown in Fig.~\ref{fig:error_etsimation_framework}, we first binarize the continuous pixel-wise error using the mean of the error distribution $\epsilon$ as the threshold and then train a UNET~\cite{ronneberger2015u} to perform classification. For the task with ROI defined, our goal is to train a classification network to output the pixel-wise probability (we refer to this probability as \emph{estimated error} in the following text):
\begin{align*}
P(|F_{ROI}(I_{HR}^{(i,j)})-F_{ROI}(I_{SR}^{(i,j)})| > \epsilon),
\end{align*}
where the superscript $(i,j)$ indexes positions. 
For the task without ROI defined, our goal is to predict the probability:
\begin{align*}
P(|I_{HR}^{(i,j)}-I_{SR}^{(i,j)}|  > \epsilon).
\end{align*}

Our preliminary experiments show significant improvement in error estimation compared to directly regressing the error. This is because most pixels in an image have very low error, significantly biasing the model to output low values. 

\subsection{Diversified Sampling}
\label{section: DPP Sampling}
After the estimated error is acquired, one naive strategy is to rescan $K$ pixels with the highest estimated error. However, due to strong correlation between neighboring pixels, 
a more cost-effective way is to sample pixels according to both saliency (quantified by estimated error) and spatial diversity. 
To this end, we formulate the problem as a determinantal point process (DPP)~\cite{kulesza2012determinantal}.

Moreover, another key contribution in our work is to propose a weighted DPP and construct a proper DPP kernel $L$ balancing saliency and spatial diversity. 
Specifically, given an image with size of $M \times M$, we construct an $N \times N$ kernel $L = U^\gamma S U^\gamma$, where $N = M^2$ is the total number of pixels in the image, $S$ is a $N \times N$ symmetric matrix indicating location similarity.
$U$ is an $N \times N$ diagonal matrix; each diagonal entry $u_{ii}$ indicates pixel $i$'s saliency, which is quantified by the estimated error described in Section~\ref{section: error estimation}. The exponent, $\gamma$, controls the trade-off between saliency and spatial diversity.  
For pixel $i$ and $j$, we have
\begin{align}
L_{ij} = u_{ii}^\gamma S_{ij} u_{jj}^\gamma,\;\;\;\;\;\;\;
S_{ij} = e^{-[(x_i-x_j)^2+(y_i-y_j)^2]/\sigma_s^2}, \label{eq:wdpp_element}
\end{align}
where $\sigma_s$ is a hyperparameter. With this new diversified sampling algorithm, our pipeline can select $K$ pixels simultaneously for rescan, while guaranteeing efficiency. The algorithm is shown in Algorithm~\ref{alg: dpp}. \textcolor{black}{In brief, conventional DPP sampling promises the diversity of sampled points for each iteration; in the current iteration, DPP finds a point which is diverse from all previous points. In contrast, our proposed WDPP finds a point which strikes a balance between diversity and saliency in each iteration.}

\begin{algorithm}
\caption{WDPP Sampling}
\SetAlgoLined
\textcolor{black}{\textbf{Input:} Location similarity matrix $S$ and diagonalized quality matrix $U$.\\
Construct kernel matrix $L = U^\gamma S U^\gamma$.\\
Compute the eigen-decomposition ${(v_n,\lambda_n)}^N_{n=1}$ of $L$.\\}
 $J \gets \emptyset$.\\
\For{$n=1,2,...,N$}{
$J \gets J \cup \{n\} $ with prob. $\frac{\lambda_n}{\lambda_n + 1}$.\\}
 $V \gets \{v_n\}_{n \in J}, Y \gets \emptyset$.\\
 \While{$|V| >0 $}{
 Select $i$ from $y$ with $Pr(i)=\frac{1}{|V|}\sum_{v\in V} (v^\top e_i)^2$.\\
 $Y \gets Y \cup i $.\\
 $V \gets V_\bot$, an orthonormal basis for the subspace of V orthogonal to $e_i$.\\
 }
\textbf{Output:} $Y$.\\
\label{alg: dpp}
\end{algorithm}

%% file: source/results.tex
\section{Experiments}
In this section, we provide an in-depth analysis of all components in our pipeline.
We use two real-world connectomics datasets, \emph{SNEMI3D} from a mouse cortex (with a resolution of $3\times3\times30$ nm/pixel), and \emph{Human} from a human cerebrum (with a resolution of $4\times4\times30$ nm/pixel), to evaluate our algorithm. 
We generate low-resolution images from original images using nearest neighbor with down-sampling rates of $\times 4$, $\times 8$, and $\times 16$. 

\subsection{Reconstruction and ROI Detection}
We reconstruct $I_{SR}$ from $I_{LR}$ using a UNET. 
Specifically, we explore two variants of UNET, one trained with an adversarial loss plus an L1 loss (SRGAN)~\cite{isola2017image} and one trained with only an L1 loss (SRUNET)~\cite{ronneberger2015u}.
Interestingly, we find that whether the adversarial loss improves reconstruction quality depends on $I_{HR}$'s noise level (see results in Supplementary). Note that reconstruction alone does not guarantee high-quality output, which is why we need adaptive rescan after binarized error estimation and diversified sampling.

For ROI detection, the ROI detector $F_{ROI}$ also uses a UNET. It is trained with $I_{HR}$ as input and human-annotated membrane as ground-truth labels.

\begin{wrapfigure}{R}{0.3\textwidth}
\centering

\includegraphics[width=0.3\textwidth]{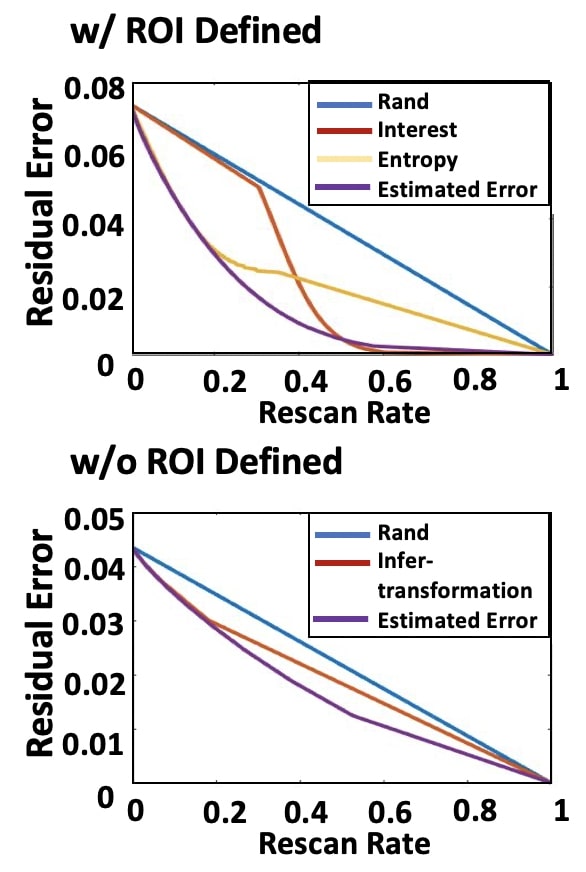}
\captionsetup{font={scriptsize}}
\caption{Sparsification error curves for estimated error (from our method) and baselines on \emph{SNEMI3D}. }
\label{fig:order correlation} 
\end{wrapfigure}

\begin{table}[t]
\setlength{\tabcolsep}{4pt}
\begin{center}
\scriptsize

\begin{tabular}{c|c|c|ccc|c|ccc}
\hline
Model &DS rate &Task &\textbf{Ours} &Entropy &Interest &Task &\textbf{Ours} &Trans &Gradient\\
\hline
\multirow{3}{*}{Bicubic} &$\times$ 4 &\multirow{3}{*}{w/ ROI} &\textbf{0.549} &0.420 &0.256 &\multirow{3}{*}{w/o ROI} &\textbf{0.488} &-- &0.323\\

&$\times$ 8 & &\textbf{0.460} &0.306 &0.329 & &\textbf{0.437} &-- &0.267\\

&$\times$ 16 & &\textbf{0.307} &0.249 &0.213 & &\textbf{0.349} &-- &0.200\\
\hline
\multirow{3}{*}{SRGAN}&$\times$ 4 & \multirow{3}{*}{w/ ROI}  &0.367 &\textbf{0.402} &0.252 &\multirow{3}{*}{w/o ROI} &\textbf{0.599} &0.451 &0.347\\
&$\times$ 8 &  &\textbf{0.495} &0.262 &0.294 & &\textbf{0.514} &0.351 &0.215\\
&$\times$ 16 &  &\textbf{0.389} &0.111 &0.317 & &\textbf{0.461} &0.293 &0.136\\
\hline
\multirow{3}{*}{SRUNET}&$\times$ 4 &\multirow{3}{*}{w/ ROI} &\textbf{0.448} &0.416 &0.238 &\multirow{3}{*}{w/o ROI} &\textbf{0.451} &0.399 &0.324\\
&$\times$ 8 & &\textbf{0.382} &0.374 &0.228 & &\textbf{0.405} &0.311 &0.250\\
&$\times$ 16 & &\textbf{0.412} &0.163 &0.046 & &\textbf{0.315} &0.203 &0.187\\
\hline

\end{tabular}
\end{center}
\caption{Correlation between ground truth error and error estimated from our method on \emph{SNEMI3D}. Baselines are interest, entropy, infer-transformation, and gradient.}

\label{Tab:error correlation}
\end{table}

\subsection{Error Estimation Analysis}

To guide the rescan process using estimated error (proposed in Section~\ref{section: error estimation}) is an effective and efficient way for active acquisition. For the first task without ROI defined, the residual error after reconstruction is $\mid I_{HR}-I_{SR} \mid$, and we use gradient~\cite{dahmen2016feature} and infer-transformation for uncertainty estimation~\cite{mi2019training,wang2018test} as baselines. For the second task with ROI defined, the residual error after reconstruction is $\mid F_{ROI}(I_{HR})-F_{ROI}(I_{SR})\mid$. We use $F_{ROI}(I_{SR})$ (interest) and its corresponding entropy as baseline. 

The first metric is the pixel-wise correlation between the estimated error and ground-truth error. 
As shown in Table~\ref{Tab:error correlation}, our method can estimate error much more accurately than the baselines (visualizations in Supplementary). This method is also robust to different reconstruction methods and down-sampling rates. The second metric is sparsification error curve~\cite{ilg2018uncertainty}, which shows how residual error decays as SEM increases the rescan rate.
Specifically, we select the top $K$ pixels according to different measurements, e.g., estimated error (our method) and entropy, and set the corresponding ground truth errors of these top $K$ pixels to zero. We then study the decay of error when $K$ increases. Fig.~\ref{fig:order correlation} shows the results for different methods including random sampling~\cite{eldar2012compressed,gan2007block}. Our method achieves the fastest decay. 

\subsection{Sampling with Saliency \& Diversity}

\begin{figure*}[t]
\begin{center}
    \includegraphics[width=1\textwidth]{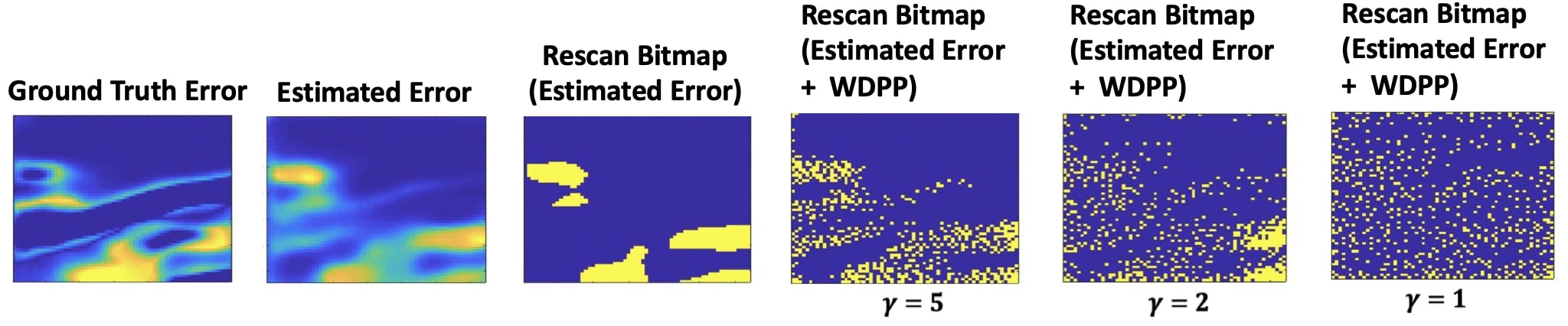}
\end{center}
   \caption{The effect of $\gamma$ to balance the trade-off between spatial diversity and saliency for WDPP sampling. All rescan bitmaps contain the same number of samples.}
\label{fig:DPP sampling with gamma} 
\end{figure*}

\begin{figure*}[t]
\begin{center}
    \includegraphics[width=1\textwidth]{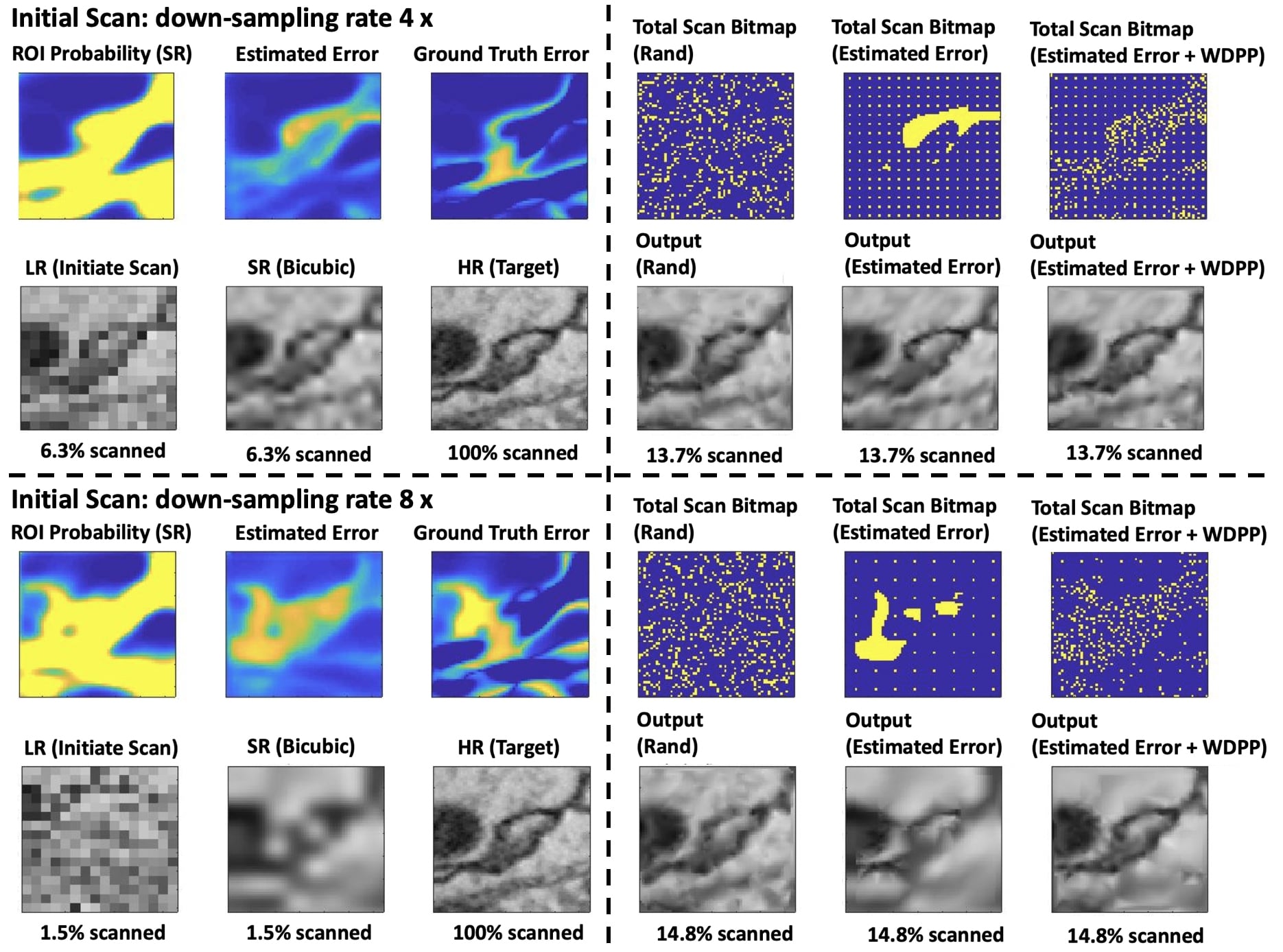}
\end{center}
\caption{The final outputs after rescan on \emph{SNEMI3D}. We compare WDPP sampling based on estimated error maps with other baselines, using down-sampling rates of $\times 4$ and $\times 8$ for initial scan. The right part shows different total scan bitmaps and their corresponding outputs using different methods. The total scan bitmaps contain the pixel locations scanned in both initial scan and rescan. }
\label{fig: image quality with DPP sampling} 
\end{figure*}

With the error estimation $U$, we can then construct the kernel matrix $L$ according to Equation~(\ref{eq:wdpp_element}) and run the WDPP sampling algorithm to select pixels to rescan. Fig.~\ref{fig:DPP sampling with gamma} shows the rescan bitmap produced by WDPP when $\gamma\in\{1,2,5\}$. As expected, (1) WDPP can naturally trade off saliency and spatial diversity during sampling; (2) compared to using only error estimation as rescan bitmaps, WDPP can cover larger areas within a sampling budget. Column 4 to 6 in Fig.~\ref{fig: image quality with DPP sampling} show the final outputs from SEM following different rescan schemes, i.e., random, estimated error, and estimated error with WDPP, demonstrating WDPP can significantly improve output image quality given a fixed rescan budget.

%% file: source/performance.tex
\section{Performance Evaluation}
In this section, we evaluate the overall speedup using our active acquisition pipeline, as shown in Fig.~\ref{fig:pipeline}. \textcolor{black}{The quality of the final output $I_{OUT}$ is compared with $I_{HR}$. For the task with ROI defined, the residual error is quantified as $ \mid F_{ROI}(I_{HR}) - F_{ROI}(I_{OUT}) \mid$. For the task without ROI defined, the residual error is quantified as $ \mid I_{HR} - I_{OUT} \mid$ (evaluations using PSNR and SSIM are shown in Supplementary).} We evaluate the initial scan with a down-sampling ratio of $\times 4$; we use different total scan rates (initial scan plus rescan) inversely proportional to the speedup factor, since the total run time of our computational pipeline on a single GPU is a small fraction ($< 3\%$) of the SEM imaging time. 
The results demonstrate that our pipeline can achieve a speedup factor of up to an order of magnitude with relatively small error.

%% file: source/conclusion.tex
\section{Conclusion}
We propose a novel and efficient learning-guided sampling algorithm based on learned saliency and spatial diversity. Our active acquisition pipeline demonstrates a potential speedup rate of up to an order of magnitude for SEM in connectomic data collection. In a broader sense, our work addresses research issues across many fields where high-throughput SEM is an essential tool for discovery. Techniques we propose in this work may also be widely used to speed up other imaging systems where sparse scanning can be applied. 

%% file: source/Acknowledgement.tex
\subsubsection{Acknowledgement}
We would thank Remco Schoenmakers, Pengfei Guo for insightful comments and suggestions. This is an area of active investigation in our laboratory and others (Uri Manor, personal communication). This research was supported by the National Science Foundation (NSF) under grants IIS-1607189, CCF-1563880, IOS-1452593 and NSF 1806818.

%% file: source/supp.tex
\section*{Appendix}

\begin{table}
\begin{center}
\small
\begin{tabular}{c|c|c|ccc|c|ccc}
\hline
Dataset &Method &Task &$\times$ 4 &$\times$ 8 &$\times$ 16 &Task &$\times$ 4 &$\times$ 8 &$\times$ 16\\
\hline
\multirow{3}{*}{SNEMI3D} &Bicubic &\multirow{3}{*}{w/ ROI} &0.089 &0.195 &\textbf{0.288} &\multirow{3}{*}{w/o ROI} &0.047 &0.076 &0.103\\
&SRGAN & &0.088 &0.212 &0.344 & &0.058 &0.091 &0.124\\
&SRUNET & &\textbf{0.074} &\textbf{0.165} &0.335 & &\textbf{0.043} &\textbf{0.069} &\textbf{0.092}\\
\hline
\multirow{3}{*}{Human} &Bicubic &\multirow{3}{*}{w/ ROI} 
&0.510 &0.700 &0.726 &\multirow{3}{*}{w/o ROI} &0.146 &0.167 &0.199\\
&SRGAN & &\textbf{0.064} &\textbf{0.111} &\textbf{0.220} & &0.166 &0.191 &0.225\\
&SRUNET & &0.664 &0.712 &0.725 & &\textbf{0.134} &\textbf{0.154} &\textbf{0.184}\\
\hline
\end{tabular}
\end{center}
\caption{The residual error after reconstruction. For the task with ROI defined, the residual error is quantified as $\mid F_{ROI}(I_{HR})-F_{ROI}(I_{SR}) \mid$. For the task without ROI defined, residual error is quantified as $\mid I_{HR}-I_{SR} \mid$. Results are evaluated on different reconstruction methods, Bicubic, SRUNET, and SRGAN on \emph{SNEMI3D} and \emph{Human}.}
\label{Tab:error after reconstruction}
\end{table}

\begin{figure}
\begin{center}
    \includegraphics[width=1\textwidth]{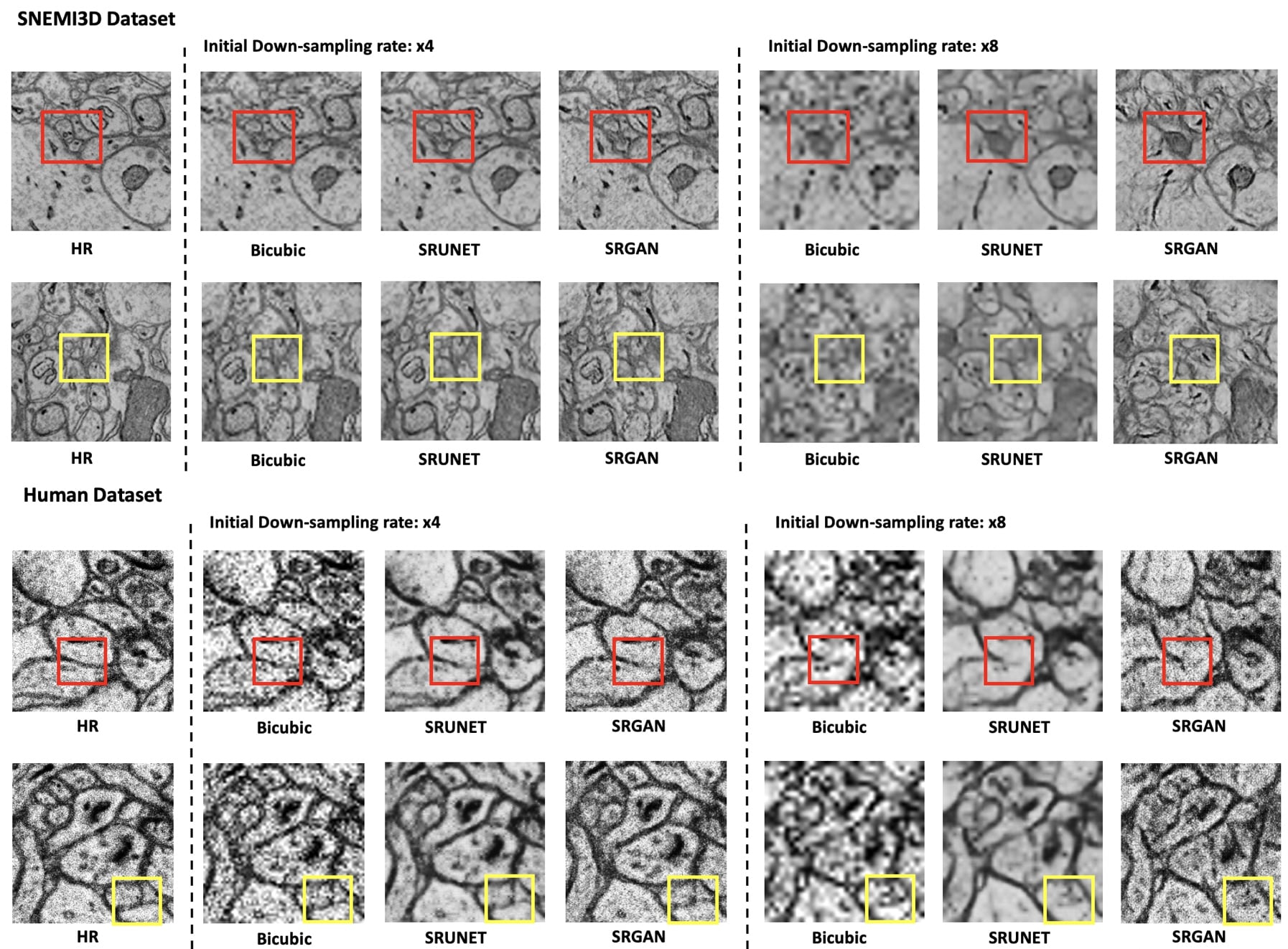}
\end{center}
\caption{The visualizations of reconstruction image quality for $I_{SR}$. Different reconstruction methods, Bicubic, SRGAN, and SRUNET are evaluated on \emph{SNEMI3D} and \emph{Human}. Adding adversarial loss (SRGAN) improves the reconstruction quality on \emph{Human} (see details in Table~\ref{Tab:error after reconstruction}), where $I_{HR}$ is relatively noisy. The red and yellow boxes mark the artifacts after reconstruction.}
\label{fig: image quality after reconstruction} 
\end{figure}

\begin{figure}[t]
\begin{center}
    \includegraphics[width=1\textwidth]{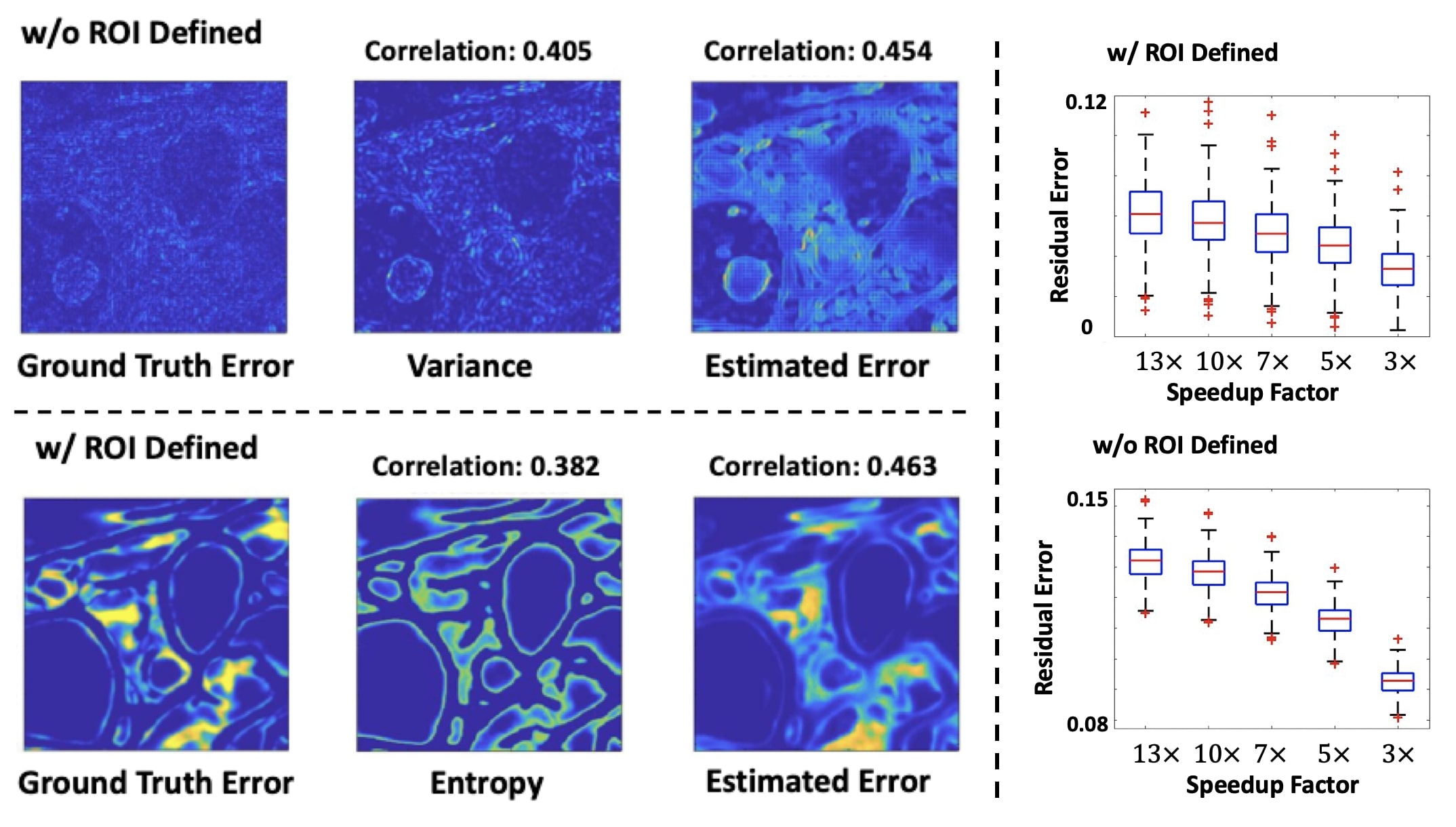}
\end{center}
\caption{\textbf{Left}: The visualizations of estimated error from our method, compared with other baselines: interest, entropy, and variance from infer-transformation on \emph{SNEMI3D}. The value of pixel-wise correlation is also identified. \textbf{Right:} The box plots represent different speedup factors v.s. residual error ($L_1$ loss) using our active acquisition pipelines on \emph{Human}. The residual error of final output $I_{OUT}$ is quantified as $\mid F_{ROI}(I_{HR})-F_{ROI}(I_{OUT}) \mid$ for the task with ROI defined, and as $ \mid I_{HR}-I_{OUT} \mid$ for the task without ROI defined.}
\label{fig:error predict quality} 
\end{figure}

\begin{table*}
\begin{center}
\small
\begin{tabular}{c|c|c|ccc}
\hline
Task &Model &Method &$\times$ 4 &$\times$ 8 &$\times$ 16\\ 
\hline
\multirow{3}{*}{w/ ROI} &\multirow{3}{*}{SRGAN} &Ours  &\textbf{0.596} &0.449 &\textbf{0.351}\\
& &Entropy &0.552 &\textbf{0.469} &0.255 \\
&  &Interest &0.255 &0.234 &0.190\\
\hline
Task &Model &Method &$\times$ 4 &$\times$ 8 &$\times$ 16\\
\hline
\multirow{3}{*}{w/o ROI} &\multirow{3}{*}{SRUNET} &Ours &0.184 &\textbf{0.179} &\textbf{0.151}\\
& &trans &\textbf{0.201} &0.158 &0.126\\
& &Gradient &0.098 &0.126 &0.107\\
\hline
\end{tabular}
\end{center}
\caption{Pixel-wise correlation between ground-truth error and error estimated from our method on \emph{Human}. Our results are compared with other baselines, interest, entropy, variance from infer-transformation, and gradient.}

\label{Tab:error correlation human}
\end{table*}

\begin{table}[t]
\begin{center}
\small
\begin{tabular}{c|c|c|c|c|c}
\hline
Speedup & 3 $\times$ & 5  $\times$ & 7  $\times$ & 10 $\times$ & 13  $\times$\\
\hline
$L_1$ Loss &0.021 &0.030 &0.035 &0.039 &0.042\\
SSIM &0.815 &0.751 &0.716 &0.686 &0.668\\
PSNR &29.48 &27.26 &26.16 &25.21 &24.57\\
\hline
\end{tabular}
\end{center}
\caption{Speedup v.s. quality for the task without ROI defined on SNEMI3D. The quality of final output $I_{OUT}$ is compared with $I_{HR}$ and evaluated with $L_1$ Loss, SSIM and PSNR.}
\label{Tab:residual error evaluated using SSIM and PSNR}
\end{table}

%% file: main.bbl
\begin{thebibliography}{10}
\providecommand{\url}[1]{\texttt{#1}}
\providecommand{\urlprefix}{URL }
\providecommand{\doi}[1]{https://doi.org/#1}

\bibitem{anderson2013sparse}
Anderson, H.S., Ilic-Helms, J., Rohrer, B., Wheeler, J., Larson, K.: Sparse
  imaging for fast electron microscopy. In: Computational Imaging XI.
  vol.~8657, p. 86570C. International Society for Optics and Photonics (2013)

\bibitem{buchholz2019content}
Buchholz, T.O., Krull, A., Shahidi, R., Pigino, G., J{\'e}kely, G., Jug, F.:
  Content-aware image restoration for electron microscopy. In: Methods in cell
  biology, vol.~152, pp. 277--289. Elsevier (2019)

\bibitem{dahmen2016feature}
Dahmen, T., Engstler, M., Pauly, C., Trampert, P., De~Jonge, N., M{\"u}cklich,
  F., Slusallek, P.: Feature adaptive sampling for scanning electron
  microscopy. Scientific reports  \textbf{6},  25350 (2016)

\bibitem{eberle2015high}
Eberle, A., Mikula, S., Schalek, R., Lichtman, J., Tate, M.K., Zeidler, D.:
  High-resolution, high-throughput imaging with a multibeam scanning electron
  microscope. Journal of microscopy  \textbf{259}(2),  114--120 (2015)

\bibitem{eldar2012compressed}
Eldar, Y.C., Kutyniok, G.: Compressed sensing: theory and applications.
  Cambridge university press (2012)

\bibitem{fang2019deep}
Fang, L., Monroe, F., Novak, S.W., Kirk, L., Schiavon, C.R., Seungyoon, B.Y.,
  Zhang, T., Wu, M., Kastner, K., Kubota, Y., et~al.: Deep learning-based
  point-scanning super-resolution imaging. bioRxiv p. 740548 (2019)

\bibitem{flegler1997scanning}
Flegler, S.L., Flegler, S.L.: Scanning \& Transmission Electron Microscopy.
  Oxford University Press (1997)

\bibitem{gan2007block}
Gan, L.: Block compressed sensing of natural images. In: 2007 15th
  International conference on digital signal processing. pp. 403--406. IEEE
  (2007)

\bibitem{helmstaedter2011high}
Helmstaedter, M., Briggman, K.L., Denk, W.: High-accuracy neurite
  reconstruction for high-throughput neuroanatomy. Nature neuroscience
  \textbf{14}(8),  1081--1088 (2011)

\bibitem{ilg2018uncertainty}
Ilg, E., Cicek, O., Galesso, S., Klein, A., Makansi, O., Hutter, F., Brox, T.:
  Uncertainty estimates and multi-hypotheses networks for optical flow. In:
  Proceedings of the European Conference on Computer Vision (ECCV). pp.
  652--667 (2018)

\bibitem{isola2017image}
Isola, P., Zhu, J.Y., Zhou, T., Efros, A.A.: Image-to-image translation with
  conditional adversarial networks. In: Proceedings of the IEEE conference on
  computer vision and pattern recognition. pp. 1125--1134 (2017)

\bibitem{46992}
Januszewski, M., Kornfeld, J., Li, P.H., Pope, A., Blakely, T., Lindsey, L.,
  Maitin-Shepard, J.B., Tyka, M., Denk, W., Jain, V.: High-precision automated
  reconstruction of neurons with flood-filling networks. Nature Methods
  \textbf{15},  605--610 (2018),
  \url{https://www.nature.com/articles/s41592-018-0049-4}

\bibitem{jarrell2012connectome}
Jarrell, T.A., Wang, Y., Bloniarz, A.E., Brittin, C.A., Xu, M., Thomson, J.N.,
  Albertson, D.G., Hall, D.H., Emmons, S.W.: The connectome of a
  decision-making neural network. Science  \textbf{337}(6093),  437--444 (2012)

\bibitem{kasthuri2015saturated}
Kasthuri, N., Hayworth, K.J., Berger, D.R., Schalek, R.L., Conchello, J.A.,
  Knowles-Barley, S., Lee, D., V{\'a}zquez-Reina, A., Kaynig, V., Jones, T.R.,
  Roberts, M., Morgan, J.L., Tapia, J.C., Seung, S., Roncal, W.G., Vogelstein,
  J.T., Burns, R., Sussman, D.L., Priebe, C.E., Pfister, H., Lichtman, J.W.:
  Saturated reconstruction of a volume of neocortex. Cell  \textbf{162}(3),
  648--661 (2015)

\bibitem{kulesza2012determinantal}
Kulesza, A., Taskar, B., et~al.: Determinantal point processes for machine
  learning. Foundations and Trends{\textregistered} in Machine Learning
  \textbf{5}(2--3),  123--286 (2012)

\bibitem{ledig2017photo}
Ledig, C., Theis, L., Husz{\'a}r, F., Caballero, J., Cunningham, A., Acosta,
  A., Aitken, A., Tejani, A., Totz, J., Wang, Z., et~al.: Photo-realistic
  single image super-resolution using a generative adversarial network. In:
  Proceedings of the IEEE conference on computer vision and pattern
  recognition. pp. 4681--4690 (2017)

\bibitem{lichtman2014big}
Lichtman, J.W., Pfister, H., Shavit, N.: The big data challenges of
  connectomics. Nature neuroscience  \textbf{17}(11),  1448--1454 (2014)

\bibitem{meirovitch2019cross}
Meirovitch, Y., Mi, L., Saribekyan, H., Matveev, A., Rolnick, D., Shavit, N.:
  Cross-classification clustering: An efficient multi-object tracking technique
  for 3-d instance segmentation in connectomics. In: Proceedings of the IEEE
  Conference on Computer Vision and Pattern Recognition. pp. 8425--8435 (2019)

\bibitem{mi2019training}
Mi, L., Wang, H., Tian, Y., Shavit, N.: Training-free uncertainty estimation
  for neural networks. arXiv preprint arXiv:1910.04858  (2019)

\bibitem{semreview}
Mohammed, A.: Scanning electron microscopy (sem): A review  (12 2018)

\bibitem{newell2016detection}
Newell, T., Tillotson, B., Pearl, H., Miller, A.: Detection of electrical
  defects with semvision in semiconductor production mode manufacturing. In:
  2016 27th Annual SEMI Advanced Semiconductor Manufacturing Conference (ASMC).
  pp. 151--156. IEEE (2016)

\bibitem{pandey2003material}
Pandey, K., Setua, D., Mathur, G.: Material behaviour: Fracture topography of
  rubber surfaces: An sem study. Polymer testing  \textbf{22}(3),  353--359
  (2003)

\bibitem{potocek2020sparse}
Potocek, P., Trampert, P., Peemen, M., Schoenmakers, R., Dahmen, T.: Sparse
  scanning electron microscopy data acquisition and deep neural networks for
  automated segmentation in connectomics. Microscopy and Microanalysis pp.
  1--10 (2020)

\bibitem{ronneberger2015u}
Ronneberger, O., Fischer, P., Brox, T.: U-net: Convolutional networks for
  biomedical image segmentation. In: International Conference on Medical image
  computing and computer-assisted intervention. pp. 234--241. Springer (2015)

\bibitem{thorpe1996speed}
Thorpe, S., Fize, D., Marlot, C.: Speed of processing in the human visual
  system. nature  \textbf{381}(6582), ~520 (1996)

\bibitem{wang2018test}
Wang, G., Li, W., Aertsen, M., Deprest, J., Ourselin, S., Vercauteren, T.:
  Test-time augmentation with uncertainty estimation for deep learning-based
  medical image segmentation. arXiv preprint arXiv:1807.07356  (2018)

\bibitem{wang2019deep}
Wang, H., Rivenson, Y., Jin, Y., Wei, Z., Gao, R., G{\"u}nayd{\i}n, H.,
  Bentolila, L.A., Kural, C., Ozcan, A.: Deep learning enables cross-modality
  super-resolution in fluorescence microscopy. Nat. Methods  \textbf{16},
  103--110 (2019)

\bibitem{weigert2018content}
Weigert, M., Schmidt, U., Boothe, T., M{\"u}ller, A., Dibrov, A., Jain, A.,
  Wilhelm, B., Schmidt, D., Broaddus, C., Culley, S., et~al.: Content-aware
  image restoration: pushing the limits of fluorescence microscopy. Nature
  methods  \textbf{15}(12), ~1090 (2018)

\bibitem{yan2017network}
Yan, G., V{\'e}rtes, P.E., Towlson, E.K., Chew, Y.L., Walker, D.S., Schafer,
  W.R., Barab{\'a}si, A.L.: Network control principles predict neuron function
  in the caenorhabditis elegans connectome. Nature  \textbf{550}(7677), ~519
  (2017)

\end{thebibliography}
